\documentstyle[preprint,aps,floats,epsfig]{revtex}
\tighten
\begin{document}
\title{Generalized scaling of the transverse mass spectrum at the Relativistic
Heavy-Ion Collider}

\author{J\"urgen Schaffner-Bielich$^a$, Dmitri Kharzeev$^b$, 
Larry McLerran$^b$, and Raju Venugopalan$^{b,c}$} 

\address{
$^a$Department of Physics, Columbia University, 538 West 120th Street, New
York, NY 10027, USA\\
$^b$Department of Physics, Brookhaven National Laboratory, 
Upton, NY 11973-5000, USA\\
$^c$RIKEN BNL Research Center, Brookhaven National Laboratory, 
Upton, NY 11973-5000, USA}
\date{\today}

\maketitle   


\begin{abstract} 
We argue that the transverse mass spectra of identified hadrons as
measured in gold-gold collisions at BNL's Relativistic Heavy-Ion Collider
(RHIC) follows a generalized scaling law. Such a scaling behavior 
is motivated by the idea of a Color Glass
Condensate, or more generally, saturation of the gluon density. 
In particular, we describe the shapes of transverse mass spectra as a function
of centrality.  This scaling of the transverse mass spectrum is shown 
to be consistent with previously observed scaling of 
multiplicity with centrality.
\end{abstract}

\pacs{}


\section{Introduction}

In a previous work, two of us argued that the mean transverse momenta
measured in relativistic heavy-ion collisions can be described by the
intrinsic transverse momentum broadening seen at the
Tevatron for $p\bar p$ collisions \cite{mlsb01}. The mean
transverse momenta of charged and identified hadrons increases
universally with the square root of the multiplicity per unit
transverse area for both $p\bar p$ and AA collisions. 
In addition, it has been shown that the total hadron multiplicity 
follows a scaling behavior motivated by the gluon saturation \cite{kn}. 

In this paper, we try to combine these two ideas in a consistent 
picture of transverse mass distributions of identified particles as a 
function of centrality. To do this, we apply theoretical ideas 
emerging in the Color Glass Condensate description of the high gluon 
density phase of QCD. Specifically, we describe 
the transverse momentum spectra using the recently measured data from
BNL's Relativistic Heavy-Ion Collider (RHIC) for AuAu collisions at
$\sqrt{s}=130$ AGeV. 

The theoretical motivation for such scaling relations is as follows. At very 
high energies, the number of partons (primarily gluons) in a nucleus 
grows very rapidly. When the occupation number of these partons is large, 
they saturate~\cite{Gribov,MuellerQiu,BlaizotMueller} and form
a novel state of matter called a Color Glass Condensate
(CGC)~\cite{kn,mv,kov,jkmw,iancu00,kmw,alex00}. 
The CGC has a bulk  
scale $Q_s$ ($\gg \Lambda_{QCD}$) which is the typical intrinsic transverse
momentum of the saturated gluons in the nucleus. The CGC can be  
probed in deeply inelastic scattering~\cite{mv2,kovmc}, in photoproduction in
peripheral heavy-ion collisions~\cite{gelis}, 
in pA collisions~\cite{km,dm01} 
and in heavy-ion collisions~\cite{kn,kmw,alex00}.
 
We remark, that scaling analysis of the transverse mass spectra at RHIC has
been also pursued in \cite{Braun} although within a different physical picture.

\section{Generalized $m_t$-scaling}

\subsection{Scaling Relations}

Collisions of heavy-ions at high energies can be 
imagined as the collision of 
two sheets of colored glass, and a large multiplicity 
of gluons is produced. 
Since the occupation number of the gluons in each of the colliding 
nuclei is large, the problem can be treated classically~\cite{kmw}. 
The only dimensionful quantities 
available to describe the collision are 
a) the transverse area of the two colliding nuclei $\sigma$ and 
b) the saturation scale $Q_s^2$ we mentioned previously, which is determined 
by the density of partons in the transverse plane.
This saturation scale is 
a function of both energy and centrality.
All dimensionful quantities can therefore be expressed in terms of powers of
one or the other scale times non--perturbative functions of the dimensionless 
product of the two scales. The initial momentum distribution of produced
gluons in high-energy collisions is
then found to be given by the relation~\cite{alex00}
\begin{equation}
        {1 \over \sigma} {{dN_g} \over {d\eta d^2p_t}} 
                        = \frac{1}{\alpha_s(Q_s^2)} 
        f_g \left(\frac{p_t^2}{Q_s^2}\right) \quad .
\label{eq:scaling_charged}
\end{equation}
Here, $f_g$ is a universal, dimensionless function for the produced gluons
which depends only on the ratio of the transverse momentum and the saturation
momentum. Different energies or scales are described by the same function with 
a correspondingly changed saturation momentum $Q_s$.

Motivated by the above relation, we want to test in this paper how well 
these scaling relations describe the actual data, i.e.\ the momentum
distribution of produced hadrons. Ab initio, it is not obvious that there is
such a relation as hadrons are produced nonperturbatively at the deconfinement
phase transition. There is also surely some transverse flow of the matter
produced in these collisions and it is not clear to what degree this flow 
might distort the distributions produced early in the CGC. Also, hard
processes are expected to modify the scaling relations at
sufficiently high transverse momenta above the saturation scale $Q_s$.

We find that the transverse momentum distributions of identified 
particles are well  described as a function only of the  
transverse mass $m_t=\sqrt{m_h^2 + p_t^2}$.  This appears to be
a good approximation so long as we are not too close to the mass of the 
particle in question.  (At small values of $p_t$ we 
expect flow effects to be important which will invalidate this simple
$m_t$ scaling form of the distributions.)
We therefore parameterize the transverse momentum spectra as
\begin{equation}
        {1 \over \sigma} {{dN_h} \over {dyd^2m_t}} = \frac{1}{\alpha_s(p_s)} 
        \kappa_h \cdot f \left(\frac{m_t}{p_s}\right) 
\quad .
\label{eq:scaling_mt}
\end{equation}
The universal function $f$ depends on $m_t$ and the corresponding saturation
scale for hadrons $p_s$ and incorporates 
$m_t$-scaling.  The constant $\kappa_h$ reflects the difference in 
abundances of various species of particles, and is of order one
for the particles which we consider.  We take it to be independent of
$p_s$. 
The hadron saturation momentum $p_s$ as well as $\sigma$ are parameters which
are determined by the energy and centrality of the collision. 
We assume $p_s$ to have the same energy 
and centrality dependence as the saturation momentum $Q_s$.
Also, the function $f$ has to be extracted from the  data. 
But once the universal function $f$ is fixed for one data set, it
should describe the  other data by rescaling the parameters $p_s$ 
and $\sigma$. Hence,
the scaling relation implies that there is one and only one function which
describes high-energy collisions at various centralities, energy and system
size. This picture can certainly be valid only in some range of these 
parameters. 
It is the purpose  of this investigation to determine
how well scaling works for data from relativistic heavy-ion collisions.

\subsection{Comparison with RHIC Data}

We turn now to a more detailed discussion of the transverse momentum
distribution focusing on identified hadron spectra as measured recently at
RHIC in AuAu collisions with $\sqrt{s}=130$ AGeV 
\cite{phenixQM,Julia01,GaborQM,phenixnew,star,starnew}. 
Note that in  relation (\ref{eq:scaling_mt}) all hadrons are described by only 
one function of the transverse mass $m_t=\sqrt{m_0^2+p_t^2}$. The transverse
mass is just the total transverse kinetic energy carried by that hadron. 
Plotting the transverse mass
spectra for identified hadrons as a function of $m_t$ (not as a function of
$m_t-m_0$) should then result in one single curve. This 
idea is not particularly
new. It is similar to the $m_t$-scaling behavior put forward by Hagedorn
for $pp$ collisions in his statistical model \cite{Hagedorn}. The universal
function in the Hagedorn model is  then a Bose-Einstein or Fermi-Dirac 
distribution and depends on one
parameter, the slope parameter $T_{\rm slope}$. 
Contrary to this picture, we approach the problem in reverse and try to
determine the universal function from the experimental data.

The transverse momentum spectra of pions, kaons, protons and antiprotons have
been reported 
by the PHENIX collaboration at RHIC to quite large transverse momenta $p_t$
\cite{Julia01}. For the first time in hadronic collisions,
one sees that the protons and antiprotons seem to be more abundant at higher
transverse momentum $p_t>2$ GeV than pions. The slope of the transverse
momentum spectra for protons and antiprotons is apparently much larger than
that of the pions, so that the nucleon spectra overshoots the pion spectra at
some $p_t>2$ GeV.

\begin{figure}[t]
\centerline{\epsfig{figure=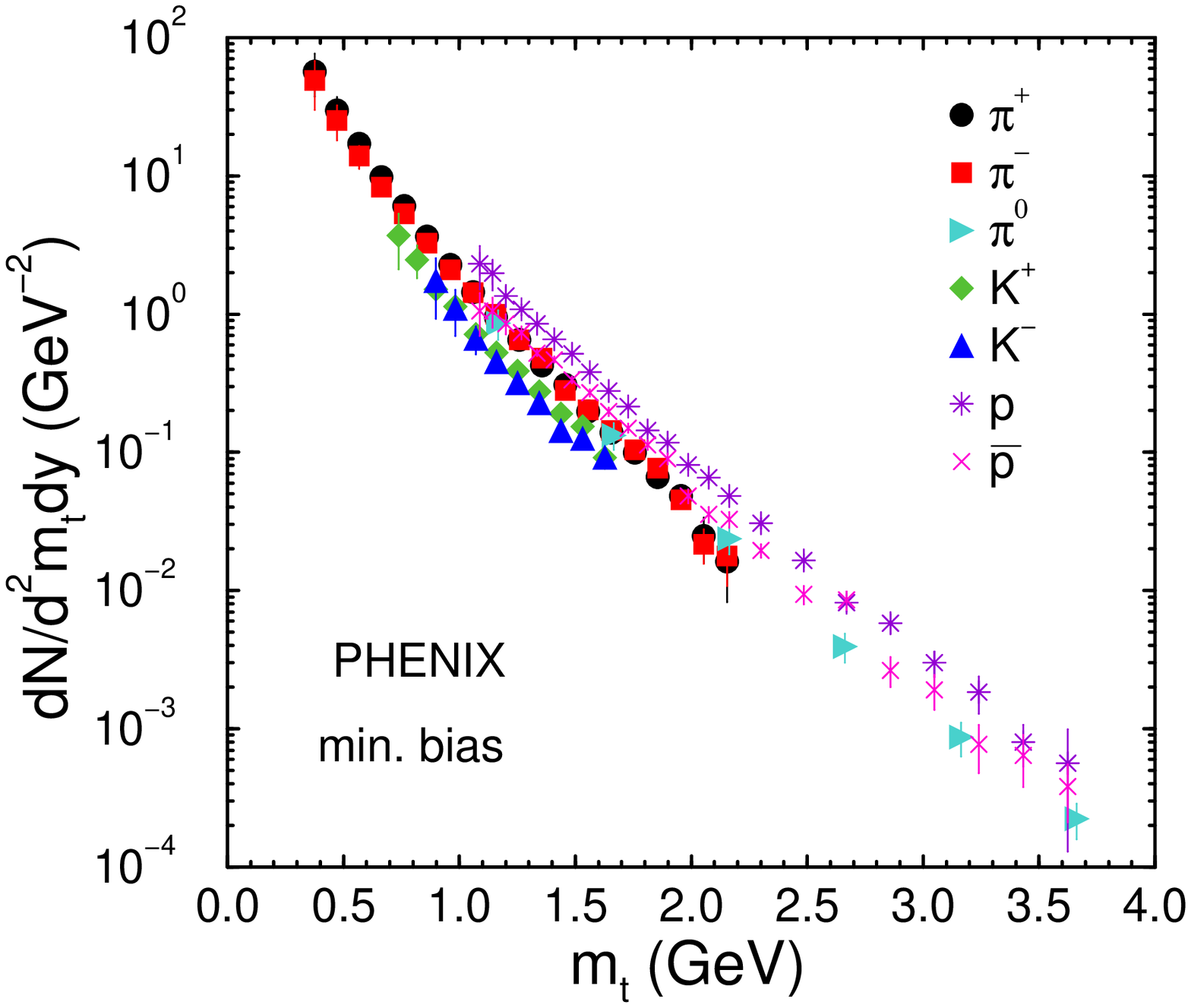,height=0.4\textheight}}
\caption{Transverse mass spectra of identified hadrons for minimum bias
gold-gold collisions as measured for $\sqrt{s}=130$ AGeV at RHIC (preliminary data taken
from \protect\cite{Julia01,GaborQM}).} 
\label{fig:mt_minbias}
\end{figure}

\begin{figure}[t]
\centerline{\epsfig{figure=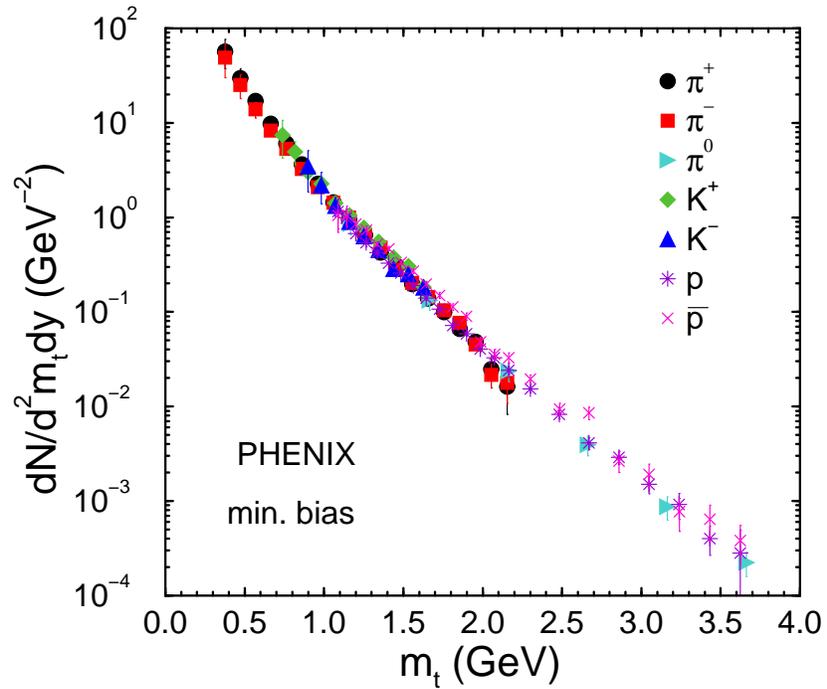,height=0.4\textheight}}
\caption{Transverse mass spectra of figure \ref{fig:mt_minbias} where the
  proton and kaon/antikaon data points are multiplied by 1/2 and 2, respectively.}
\label{fig:mt_minbias_scaled}
\end{figure}

The scaling relation (\ref{eq:scaling_mt}) can now be tested by plotting the
transverse mass spectra. 
Figure~\ref{fig:mt_minbias} shows the preliminary minimum bias data of the
PHENIX collaboration for charged pions, kaons, protons and antiprotons
\cite{Julia01} and for neutral pions \cite{GaborQM} as a function of $m_t$
(not $p_t$).  
As one can see from Fig.~\ref{fig:mt_minbias}, the data points for all 
hadrons seem to follow one curve 
over several orders of magnitude for a broad range of
transverse mass. 
The neutral pion data is on top of the antiproton data within the error bars. 
But one notices also, that the curves
for kaons and protons are shifted downwards and upwards, respectively. These
are effects due to quantum numbers (strangeness and baryon number) which are
not taken into account in the scaling relation. The production of strange
particles is suppressed due to the strange quark mass, while the number of
protons is enhanced due to the initial baryon number excess coming from the
two colliding nuclei. Interestingly the proton-antiproton pair production seems
to be not suppressed relative to pion production but kaon production is. 
We note that these changes are moderate, they shift the
curves of the kaons down by about a factor two while shifting the proton curve
up by about a factor two from the universal curve. 
These shifts correspond to the constant $\kappa_h$ in eq.~\ref{eq:scaling_mt}.
Shifting the kaon and the proton spectra by these factors results
in Fig.~\ref{fig:mt_minbias_scaled}. 
Now all the data points are lying on one curve.
It tells us that the shape of the spectra for the same value of $m_t$ are
equal for pions, kaons, and nucleons. (Often times, 
a slope parameter is extracted to characterize spectra.  This
is extracted from the spectra close
to the mass threshold, i.e.\ at $m_t=m_h$.
The larger the mass of the hadron $m_h$ the
larger will be its slope parameter $T_{\rm slope}$,
as the slope of the transverse mass spectra is increasing with $m_t$.
We can read from 
Fig.~\ref{fig:mt_minbias_scaled}, that this effect will also continue for
heavier particles but it is less pronounced as the slope 
of the curve levels off.)
If scaling works for all particles, the transverse mass distribution of
hyperons will then follow also the same universal $m_t$-curve of
Fig.~\ref{fig:mt_minbias_scaled} modulo shape independent 
effects from baryon and strangeness
number conservation which affect only the overall normalization of the
distribution.

Being more quantitative, the transverse mass spectra of
Fig.~\ref{fig:mt_minbias_scaled} can be fitted by a power law of the form 
$(1+m_t/p_s)^{-n}$ with the parameters $n=16.3$ and $p_s=2.71$ GeV. We
remark that the values for $n$ and $p_s$ are strongly correlated, fits with
different values for $p_s$ and properly adjusted $n$ gives an equally good fit 
to the data points. 
The values for $n$ and $p_s$ are constrained in such a way that they 
get about the same mean transverse mass of 
\begin{equation}
\langle m_t \rangle = \frac{2p_s}{n-3}
\end{equation} 
for each fit.
Let us define now the local slope as
\begin{equation}
-\frac{1}{T_{\rm slope}} = \frac{d}{dm_t} \ln \left(f(m_t/p_s)\right)
\quad .
\end{equation}
Then we find that the local slope parameter for a power-law distribution in
$m_t$ is given by 
\begin{equation}
T_{\rm slope} = \frac{p_s}{n} + \frac{1}{n} m_t
\label{eq:tslope}
\quad .
\end{equation}
i.e.\ a constant term plus a term linear in the transverse mass. The constant 
is closely related to the mean transverse mass. The correction term
proportional to the transverse mass originates 
from the non-exponential behavior of the
power-law distribution which enhances the high-$m_t$ part of the
distribution. With the fitted parameters given above, we find the following
slope parameters for various hadrons at $m_t=m_h$: 
175 MeV ($\pi$), 196 MeV (K), 224 MeV (p), 235 MeV ($\Lambda$), 247 MeV
($\Xi$). The actual measured values for $\pi$, $K$ and nucleons are higher
\cite{Julia01} as the experimental fits are taken 
around $0.5$ GeV above the threshold. 
Hence, the apparently different slopes for hadrons can be also
explained by a generalized $m_t$-scaling of the transverse mass spectra. 
Note, that the relation (\ref{eq:tslope}) is similar to the one derived for
radial flow of non-relativistic particles 
$T_{\rm slope} = T_0 + 0.5\beta^2\cdot m_h$
\cite{flowslopes}. 
In $pp$ collisions, the value for $p_s$ is smaller than in $AA$ collisions, so
that the deviations from an exponential shape of the transverse mass
distribution are less pronounced. Therefore, 
one recovers in $pp$ collisions
the traditional $m_t$-scaling behavior of 
Hagedorn's statistical model \cite{Hagedorn}
with about similar slope parameters for all hadrons. We will discuss the
centrality dependence below in more detail.

We remark that the universal $m_t$-scaling seen in 
Fig.~\ref{fig:mt_minbias} does not rule out radial flow because
at large momenta the hydro picture also predicts 
a universal curve for all hadrons.
A clear indication of radial flow would be a
deviation from the $m_t$-scaling curve. 
These deviations should be especially pronounced for heavier
hadrons.  
STAR measures at even lower $p_t$ than PHENIX and reports a
larger slope parameter for antiprotons than PHENIX \cite{star,starnew}. This
experimental finding cannot be explained by generalized $m_t$-scaling. 
The flatter distribution at  small $m_t$ close to $m_h$ compared to larger
$m_t$ can be well described in hydrodynamical models with collective
transverse flow \cite{pasi,srivastava,derek}. 
In general, radial flow implies a violation of $m_t$ scaling,
as the transverse momentum distribution then depends then on both, $p_t$ and 
$m_t$.   

We expect that there should be radial flow effects in heavy ion
collisions.  These should show up in a relatively unambiguous way
as the deviations from $m_t$ scaling for $m_t$ near threshold.  What we
try to describe in this paper are the gross features of the data for
$m_t$ far from threshold.  
The consistency of the picture we present is simple within the Color Glass
Condensate framework. It may turn out for reasons which we do not
understand that radial flow preserves these simple patterns.
For example, this might be plausible if the flow sets in very 
early in the collision.  It  might also turn
out that the patterns we observe have some as yet undiscovered interpretation.
The scaling relations we present nevertheless would remain as a simple
phenomenological description of the data.

Finally, we note that the universal $m_t$-scaling for hadrons in heavy-ion
collisions has been also observed at lower bombarding energies at CERN's SPS
\cite{heinz,nuxu} but it has not been discussed in detail.  
In particular, no interpretation in terms of scaling has been given.

\section{Scaling with Centrality}

The scaling relation (\ref{eq:scaling_mt}) predicts more than the universal
curve for the transverse mass spectra. As it should be valid for all
centralities, the transverse mass spectra should scale for each centrality
class. Moreover, there is only one scaling function and the transverse mass
spectra for different centralities can then be rescaled into each other by
properly choosing the transverse area $\sigma$ and the momentum $p_s$ for each 
centrality bin.

\begin{figure}[t]
\centerline{\epsfig{figure=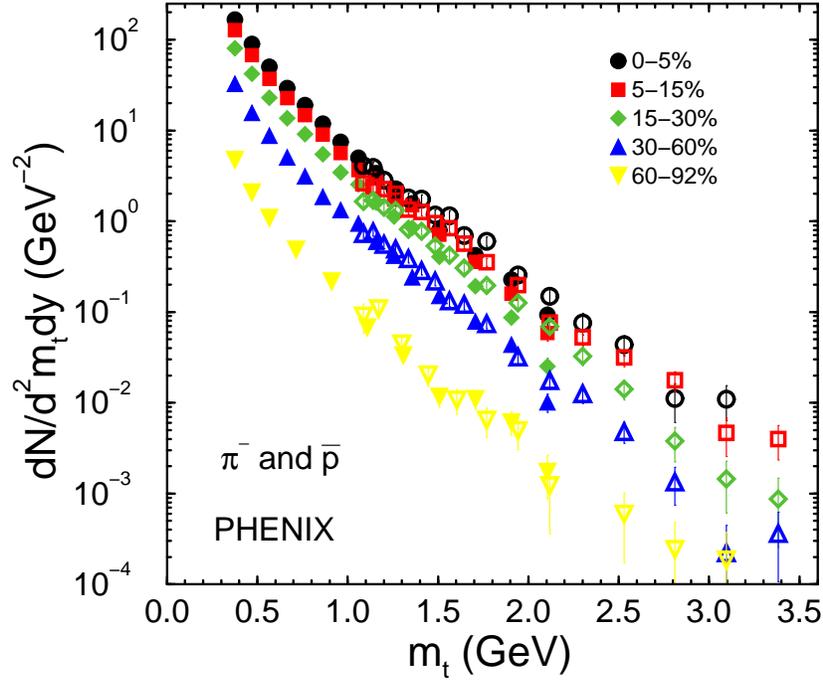,height=0.4\textheight}}
\caption{Transverse mass spectra of $\pi^-$ and $\overline{\rm p}$
for different centralities at RHIC 
(preliminary data as published in \protect\cite{Julia01}). Filled symbols are for $\pi^-$,
open symbols for antiprotons.}
\label{fig:mt_centrality}
\end{figure}

The first statement is probed in Fig.~\ref{fig:mt_centrality} which shows the
preliminary $m_t$-spectra of PHENIX for negatively charged pions and
antiprotons for different centralities \cite{Julia01}. Indeed, it seems
that there is a universal function for each centrality bin, even
for the most peripheral one, which describes the pion and antiproton spectra
simultaneously. The form of the curves looks similar for the
various centralities, too.

To check that the $m_t$ distribution is universal, we rescale the data points
in absolute normalization and in transverse mass: 
\begin{equation}
        {1 \over \sigma} {{dN_h} \over {dyd^2m_t}} \to \frac{1}{\lambda}
        {1 \over \sigma} {{dN_h} \over {dyd^2m_t}} 
\qquad \mbox{ and } \qquad
m_t \to \frac{m_t}{\lambda'} 
\end{equation}
so that the data points for all centralities are lying on top of each other. We
choose the most central bin as a reference curve.
The scaling parameters $\lambda$ and $\lambda'$ are extracted for each
centrality bin. The scaling parameter $\lambda$ is just 
$\sigma/\alpha_s(p_s)$ for a given centrality relative 
to the value for the most central bin. The
parameter $\lambda'$ is the momentum $p_s$ for a given centrality bin
divided by the one for the most central bin. Technically, we use the
$p_t$-spectra of charged particles as measured by the PHENIX collaboration
\cite{phenixQM} to fix the scaling parameters $\lambda$ and $\lambda'$ for
each centrality bin as the data has much better statistics. Then we use those
parameters to rescale the $m_t$ distributions.  

\begin{figure}[t]
\centerline{\epsfig{figure=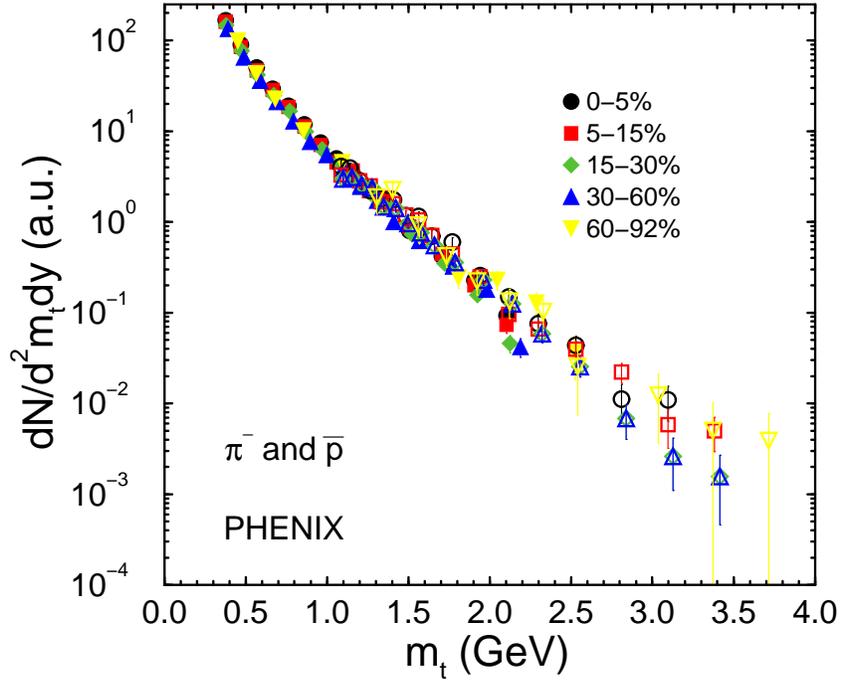,height=0.4\textheight}}
\caption{Rescaled transverse mass spectra of $\pi^-$ and $\overline{\rm p}$
of the previous figure.}
\label{fig:mt_scaling}
\end{figure}

Fig.~\ref{fig:mt_scaling} shows the rescaled version of the transverse
mass spectra for all centralities. The data points for all centralities are
now lying on top of each other, i.e.\ scaling works 
reasonably well also for the centrality
dependence of the $m_t$-spectra. 
Accordingly, also the charged particle $p_t$ spectra then can be
rescaled for each centrality cut. There seem to be deviations from scaling at
large $m_t$ in Fig.~\ref{fig:mt_scaling} as that the more central bins are
suppressed relative to the peripheral ones. But within the error bars it is
difficult to make a more definite conclusion. We see a similar trend at high
$p_t$ for the rescaled charged particle spectra,
but again within the error bars it is hard to draw a conclusion.

One expects that $\sigma/\alpha(p_s)$ rescales as a product of the transverse
area times a logarithmic function of the centrality. Similarly, one expects
$p_s^2$ to scale as the cube root of the centrality.
In the following we will check, if this is the case.

\subsection{Scaling of the Transverse Area}
\label{sec:scalingA}

\begin{figure}[t]
\centerline{\epsfig{figure=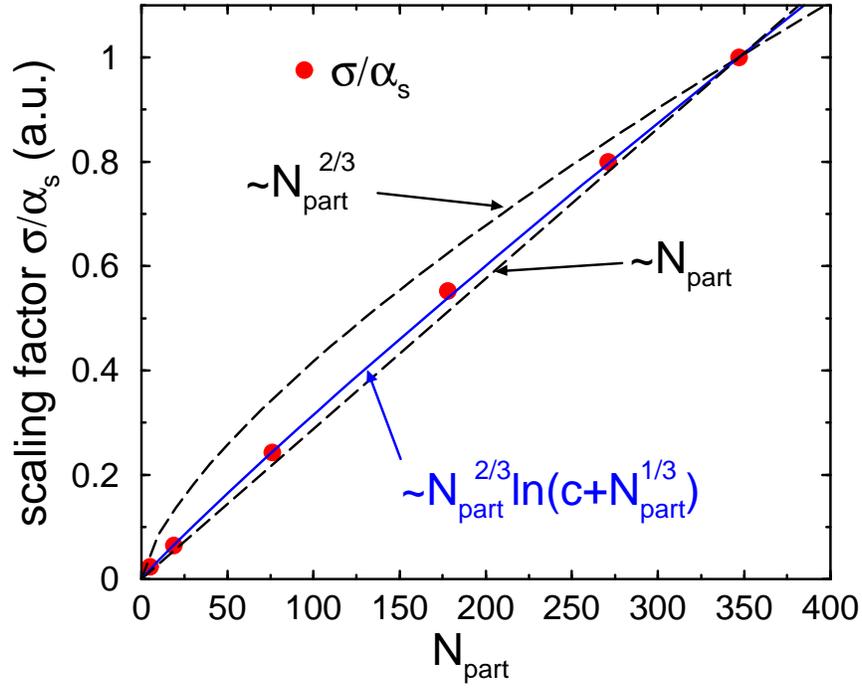,height=0.4\textheight}}
\caption{Scaling factor of the absolute normalization of the transverse mass
  spectra. All curves are normalized to the most central bin at $N_{\rm part}=347$.}
\label{fig:scaling_area}
\end{figure}

The fitted scaling factor $\sigma/\alpha_s(p_s)$ for the different centrality 
bins are plotted in Fig.~\ref{fig:scaling_area} versus the number of 
participants $N_{\rm part}$. 
We take the number 
of participants as reported by the PHENIX collaboration in \cite{phenix_prl}
for their centrality cuts. The scaling factors are given relative to the one
for the most central bin at $N_{\rm part} = 347$.

One sees from the figure, that the fitted scaling factors 
seems to have a more linear dependence on 
$N_{\rm part}$ than to follow the curve 
for a transverse area $A\sim N^{2/3}_{\rm part}$. 
Nevertheless, there is a correction
factor coming from $\alpha_s(p_s)$. 
The quantity $\alpha_s$ is taken as
$\alpha_s^{-1} \sim \ln \left(p_s^2/\Lambda_{\rm QCD}^2\right)$ and the
saturation momentum is dependent on the number of
participants $N_{\rm part}$ \cite{kn}. The factor $\alpha_s$ induces then a
logarithmic correction to the scaling factor.
The scaling factor should then follow a curve proportional to the transverse
area $A$ divided by $\alpha_s$ which is of the form 
$N_{\rm part}^{2/3}\ln\left(p_s^2/\Lambda_{\rm QCD}^2\right)$. 
For consistency, the argument
of the logarithm is taken from the dependence of $p_s$ on $N_{\rm part}$ as
will be discussed in the next section. 
The dependence of $p^2_s$ with $N_{\rm part}$ 
has been fitted to be $p_s^2/p^2_{s,c}=0.61+0.39(N_{\rm part}/347)^{1/3}$ 
(see figure \ref{fig:scaling_ps}) and $p_{s,c}$ is the momentum scale for the
most central bin. 
The dependence of $\alpha_s$ with $p_s$ is then of the form
$\alpha_s^{-1}\sim \ln\left(\left(0.61+0.39(N_{\rm
      part}/347)^{1/3}\right)/\mu^2\right)$ where $\mu^2$ is the ratio of
$\Lambda_{\rm QCD}^2$ and $p^2_{s,c}$ and adjusted to be 0.6.
The corresponding curve is 
shown in Fig.~\ref{fig:scaling_area} by the solid line and now follows very
closely the extracted values for the scaling factor.

\subsection{Scaling of the Transverse Momentum}

The second scaling factor, $p_s$, as extracted from the data is 
shown in Fig.~\ref{fig:scaling_ps}. Again, the factors are normalized to the
one for the most central bin. The centrality dependence of $p_s$ is
quite weak for moderate to large number of participants but falls off for 
peripheral collisions. 
The fitted values of $p_s$ are compared with the expected 
behavior for the saturation
momentum $Q^2_s\sim N^{1/3}_{\rm part}$ in the figure. 
It is seen, that $p_s$ changes less
rapidly with centrality than $N^{1/6}_{\rm part}$. The reason is that the
gluon densities at hand are still too small to reach this behavior fully. 
Fits of the form  $\left(c + N^{1/3}_{\rm part}\right)^{1/2}$ describe
the centrality dependence of $p_s$ much better (see figure and the previous
discussion in section \ref{sec:scalingA}).
Hence, as $p_s$ increases it appears to be tending towards 
the scaling with $N_{\rm part}^{1/6}$.

Let us compare $p_s$ to 
the mean transverse momentum defined by
\begin{equation}
\langle p_t \rangle = 
\frac{p_s^3 \cdot \int_{m_h/p_s}^\infty d^2x  \sqrt{x^2-(m_h/p_s)^2}  f(x)} 
{p_s^2 \cdot \int_{m_h/p_s}^\infty d^2x  f(x)} 
= p_s \cdot P\left(\frac{m_h}{p_s}\right) 
\label{eq:meanpt}
\quad ,
\end{equation}
where $m_h$ is the vacuum mass of the identified hadron and $P$ is some
function of the ratio of the hadron mass and the scaling momentum.
The momentum $p_s$ is proportional to the mean $p_t$. In the limit 
$N_{\rm part} \to 2$ one should recover the mean $p_t$ and its corresponding
momentum $p_s$ of $pp$ collisions. The constant seen in the fit of $p_s$ with
$N_{\rm part}$ then stands for that finite mean 
$p_t$ (or $p_s$) already present in $pp$
collisions.  
Note, that $p_s$ is not directly identical 
with the mean $p_t$ but differs by the factor
$P$. As this factor $P$ depends on the vacuum mass of the hadron,
the mean $p_t$ for each hadron follows a different behavior with
centrality which depends on the ratio of its vacuum mass to the momentum $p_s$
in a nontrivial way.  

The weak centrality dependence of $p_s$ is 
compatible with the small change of the mean $p_t$ with $N_{\rm part}$
seen at RHIC \cite{Julia01,zhangbu}. 
The mean $p_t$ for charged particles measured in
$p\bar p$ collisions at $\sqrt{s}=200$ GeV has been measured by the UA1
collaboration to be $\langle p_t \rangle = 392 \pm 3 $ MeV \cite{ua1}.
The energy dependence of the mean $p_t$ for $p\bar p$ collisions is known to
be quite weak. 
STAR reports preliminarily that $\langle p_t \rangle = 508\pm 12$ MeV for
central AuAu collisions at $\sqrt{s}=130$ GeV \cite{star,zhangbu}.  
So, the ratio of mean $p_t$ squared for $N_{\rm part} \to 2$ to the 
value for central AuAu collisions,
i.e.\ the value for $p^2_s$ at $N_{\rm part}=2$, is calculated to be 0.60 which
is in agreement with the results of Fig.~\ref{fig:scaling_ps}.

The centrality dependence for $p_s$ can be compared to the measured transverse
energy per charged particle by the PHENIX collaboration \cite{phenix_et} which 
is approximately proportional to the mean transverse momentum. The data points 
are flat from the most central bin down to $N_{\rm part}\sim 76$. 
The extracted values for $p_s$ in Fig.~\ref{fig:scaling_ps}
follow closely this constant dependence on centrality for these values of
$N_{\rm part}$.

\subsection{Multiplicity}

We can cross-check our scaling relations by computing the centrality
dependence of the charged multiplicity.
Integrating eq.~(\ref{eq:scaling_mt}) over the transverse mass gives
the multiplicity 
\begin{equation}
{{dN}\over {dy}} 
 = {\sigma \over \alpha_s} p_s^2 \cdot \kappa_h \int_{m_h/p_s}^\infty \!\!\! d^2x  f(x) 
 = {\sigma p_s^2 \over \alpha_s} \cdot \kappa_h \ 
F \left(\frac{m_h}{p_s}\right) 
\quad ,
\label{eq:dndy}
\end{equation}
The dependence of the multiplicity on the scaling factors is analogous to the
multiplicity dependence of initially produced gluons in the CGC (see e.g.\
\cite{kn} and references therein):
\begin{equation}
{{dN}\over {dy}} \sim {\sigma Q_s^2 \over \alpha_s}
\quad ,
\end{equation}
except for the last factor $\kappa_h$ which takes into account 
normalization effects 
associated with different species of particles. These are small corrections 
to these formula, as e.g.\  the kaon spectra is
lying about a factor two below the universal curve in $m_t$ (see our
discussion of Fig.~\ref{fig:mt_scaling}).
Neglecting the dependence of the shift from rapidity
$y$ to pseudo-rapidity $\eta$ on centrality, 
the charged multiplicity density reads 
\begin{equation}
\frac{1}{N_{\rm part}}{{dN}\over {d\eta}} \sim 
{\sigma p_s^2 \over \alpha_s N_{\rm part}} 
\left[\kappa_{\pi} F\left(\frac{m_\pi}{p_s}\right) + \kappa_K 
F\left(\frac{m_K}{p_s}\right)
  +\kappa_p F\left(\frac{m_p}{p_s}\right) +
  \dots \right] 
\quad .
\label{eq:multiplicity}
\end{equation}
If we take the universal function $f$ to be a 
power law in the transverse mass:
\begin{equation}
f\left(\frac{m_t}{p_s}\right) \sim
\left(1+\left(\frac{m_t}{p_s}\right)\right)^{-n} 
\end{equation}
then it can be analytically integrated to get
\begin{equation}
F\left(\frac{m_h}{p_s}\right) \sim 
\frac{1}{(n-2)(n-1)}
\left(\frac{p_s}{m_h+p_s}\right)^{n-1}
\left(1+(n-1)\frac{m_h}{p_s}\right)
\quad .
\end{equation}
Indeed, one can perform a power law fit to the transverse mass spectra of the
preliminary PHENIX data of the most central bin 
of Fig.~\ref{fig:mt_scaling} with $n=11.8$ and $p_s=1.65$ GeV. 
We can now compute the charged multiplicity from the two scaling factors 
and the power law function for the most central bin. 

Quantitatively, the factor $\sigma/\alpha_s$ grows approximately like  
$N^{2/3}_{\rm part}\ln(c+ N^{1/3}_{\rm part})$ (see
Fig.~\ref{fig:scaling_area}) and 
$p^2_s$ grows like  
$c+N^{1/3}_{\rm part}$ (see Fig.~\ref{fig:scaling_ps}), so that finally we
recover  
\begin{equation}
\frac{1}{N_{\rm part}}{{dN}\over {d\eta}} \sim 
\ln (c+N^{1/3}_{\rm part}) \left(1 + N_{\rm part}^{-1/3}\right) .
\end{equation}
This form is 
the one expected from the high density QCD
approach with a nearly constant contribution from soft
scatterings and a term growing with $\ln N_{\rm part}$ \cite{kn}.
It can be motivated by noting that at large $N_{\rm part}$ it has
a form consistent with our expectations concerning saturation.
At low $N_{\rm part}$, the constant gets important reflecting non-perturbative
physics, 
perhaps associated with a vacuum gluon condensate \cite{kl1,kkl}.  
That the dependence of the charged multiplicity on 
$N_{\rm part}$ comes out correctly is nontrivial. The multiplicity of
particles is dominated by the contributions from small $p_t$ where it is 
possible to have deviations from 
the universal function due to resonance decays. 
As it turns out this does not seem to be the case qualitatively.

\section{Transverse Momentum Broadening Revisited}

The eqs.~(\ref{eq:meanpt}) and (\ref{eq:multiplicity}) give a relation between the
mean transverse momentum and the charged multiplicity:
\begin{equation}
\langle p_t \rangle^2 \sim \alpha_s P^2\left(m_h/p_s\right) \cdot
\frac{1}{\sigma} \frac{dN}{d\eta}  
\label{eq:pt2dndeta}
\quad .
\end{equation}
Except for the factor $\alpha_s P^2$, this is the relation as studied in
\cite{mlsb01}. The mean momentum squared increases as the charged multiplicity 
per transverse area.
This has been found to be in good agreement 
with the $p\bar p$ data from the Tevatron and the
heavy-ion data from the SPS \cite{mlsb01}. We expect a similar behavior 
at RHIC from scaling arguments. 
Compared to the previous work \cite{mlsb01}, we are now in the
position to explain in more detail the effects of $p_t$ broadening, 
i.e.\ that the
mean $p_t$ rises more rapidly for more massive particles compared to the
purely kinematic behavior as seen in $p\bar p$ collisions at the Tevatron 
\cite{e735_88,e735_90,e735_93}.
The mass dependent factor $P(m_h/p_s)$ gives a different steepness in the
increase of mean $p_t$ squared for different hadrons, so that pions, kaons and
nucleons have a different slope. 
As $P(m_h/p_s)$ increases with hadron mass $m_h$, the slope
will be larger for more massive particles. If the function $f$ is of
exponential form, $f \sim \exp (-m_t/T)$, then the mean $p_t$ is given by
\begin{equation}
\langle p_t \rangle = T \left(\frac{\pi m_h}{2T}\right)^{1/2}
\frac{K_{2}(m_h/T)}{K_{3/2} (m_h/T)}
\quad .
\end{equation}
For heavy particles, $m_h\ll T$, one gets 
$\langle p_t \rangle^2 \sim \pi/2  m_h T$ which is besides a small
correction factor just the purely kinematic factor $2m_h T$ for
non-relativistic particles.

\begin{figure}[t]
\centerline{\epsfig{figure=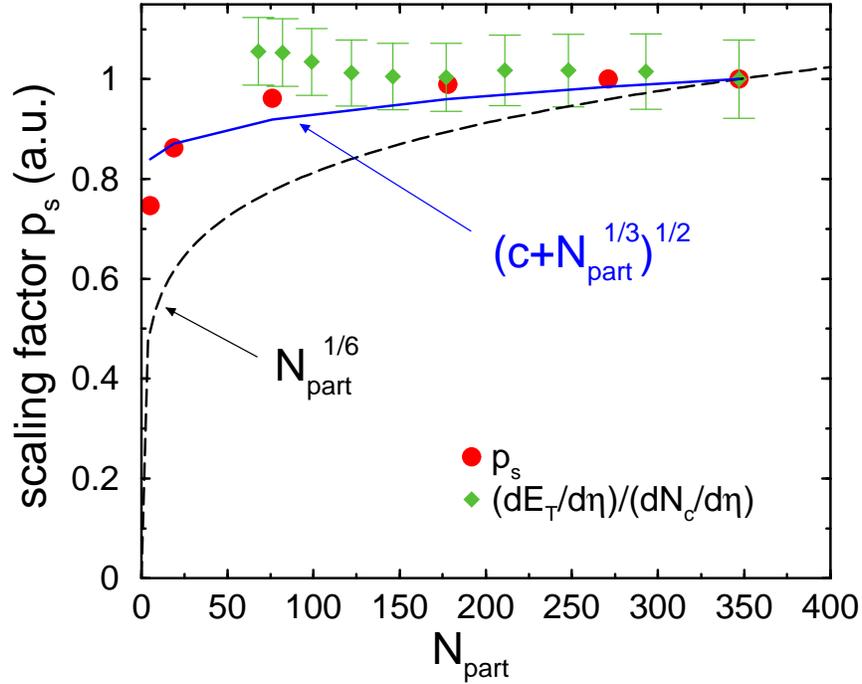,height=0.4\textheight}}
\caption{Momentum scaling factor $p_s$ as extracted from the rescaling of the
$m_t$-spectra. The data for the transverse energy per charged particle from
PHENIX \protect\cite{phenix_et} is plotted for comparison and normalized to the
most central bin.} 
\label{fig:scaling_ps}
\end{figure}

We show in Fig.~\ref{fig:scaling_ps} as a function
of centrality the PHENIX data on the transverse energy 
per charged particle $E_T/N_c$ \cite{phenix_et}. 
The data has large systematic errors, and taken alone
would provide no evidence of $p_t$ broadening. The errors on this data
nevertheless allow for consistency with the data on identified particle 
spectra. 
The STAR data on mean $p_t$ for charged particles
shows a centrality dependence which appears consistent 
with our analysis \cite{star,zhangbu}.

 
\acknowledgments

We thank Barbara Jacak, Julia Velkovska, and 
Nu Xu for many helpful discussions. 
J.S.B. acknowledges RIKEN, BNL, and the 
U.S. Department of Energy for providing the
facilities essential for the completion of this work. 
This manuscript has been authorized with the U.S. Department of Energy under
Contract No.\ DE-AC02-98CH10886 and No.\ DE-FG-02-93ER-40764.



\newpage

\end{document}